\documentclass[thmsa,11pt,letterpaper]{article}
\usepackage{amssymb}
\usepackage{amsfonts}
\usepackage{amsmath}

\input{tcilatex}

\begin{document}

\setcounter{page}{0} \topmargin0pt \oddsidemargin5mm \renewcommand{%
\thefootnote}{\fnsymbol{footnote}} \newpage \setcounter{page}{0} 
\begin{titlepage}
\begin{flushright}
EMPG-05-02 \\
\end{flushright}
\vspace{0.5cm}
\begin{center}
{\Large {\bf A Q-Operator Identity for the Correlation Functions of the Infinite XXZ Spin-Chain} }

\vspace{0.8cm}
{ \large Christian Korff}

\vspace{0.5cm}
{\em Centre for Mathematical Science, City University\\
Northampton Square, London EC1V 0HB, UK }
\end{center}
\vspace{0.2cm}
 
\renewcommand{\thefootnote}{\arabic{footnote}}
\setcounter{footnote}{0}

\begin{abstract}
An explicit construction for Q-operators of the finite XXZ spin-chain with twisted boundary conditions is presented. 
The massless and the massive regime is considered as well as the root of unity case. It is explained how these 
results yield an alternative expression for  the trace function employed in the description
of  the correlation functions of the inhomogeneous XXZ model on the infinite lattice by Boos, Jimbo, Miwa, Smirnov and Takeyama. 
\end{abstract}
\vfill{ \hspace*{-9mm}
\begin{tabular}{l}
\rule{6 cm}{0.05 mm}\\
e-mail: c.korff@city.ac.uk\\
\end{tabular}}
\end{titlepage}\newpage

\section{Introduction}

The computation of correlation functions is one of the major challenges in
integrable systems. Multidimensional integral formulae have been derived for
the infinite \cite{JMMN,JMbook,JM} as well as the finite XXZ spin-chain at
zero \cite{KMT,KMST} and at finite temperature \cite{GKS}. More recently it
has been observed that some of these multidimensional integrals can be
reduced to one-dimensional ones allowing for their explicit computation \cite%
{KSTS1,KSTS2,TKS}. In a series of papers \cite{BKS1,BKS2,BKS3} this
reducibility was connected to a duality between the solutions of the quantum
Knizhnik-Zamolodchikov (qKZ) equations of level 0 and level -4 which the
correlation functions of the model ought to obey \cite{IIJMNT}.

In this article we will refer to the subsequent work by Boos, Jimbo, Miwa,
Smirnov and Takeyama on the correlation functions of the infinite XXZ
spin-chain \cite{BJMST},%
\begin{equation}
H_{\text{XXZ}}=\frac{1}{2}\sum_{j=-\infty }^{\infty }\left\{ \sigma
_{j}^{x}\sigma _{j+1}^{x}+\sigma _{j}^{y}\sigma _{j+1}^{y}+\frac{q+q^{-1}}{2}%
\,\sigma _{j}^{z}\sigma _{j+1}^{z}\right\} ,\quad q=e^{i\pi \nu }\ .
\label{H}
\end{equation}%
In there it is explained that the correlation functions are particular
solutions of a reduced set of qKZ equations which can be formulated in terms
of a special trace of a certain monodromy matrix. The latter formally
resembles in its algebraic structure the monodromy matrix of the
inhomogeneous six-vertex model on a \emph{finite} lattice whose number of
columns is determined by the number of operators in the correlation
function. It is the special trace of this finite monodromy matrix which we
will identify as an analytic continuation of the six-vertex fusion hierarchy
in terms of $Q$-operators. Before giving the technical details of the
computation we state this result more explicitly. To this end we first
briefly summarize the main outcome of the work \cite{BJMST}.

\subsection{Correlation functions and a generalized trace}

Consider the correlation functions of the elementary matrices $%
(E_{\varepsilon _{j},\bar{\varepsilon}_{j}})_{j}$ acting on the $j^{\text{th}%
}$ site in the XXZ spin-chain and interpret 
\begin{equation}
h_{p}(\lambda _{1},...,\lambda _{p})^{\varepsilon _{1},...,\varepsilon _{p},%
\bar{\varepsilon}_{p},...,\bar{\varepsilon}_{1}}=\left\langle vac\right\vert
(E_{-\varepsilon _{1},\bar{\varepsilon}_{1}})_{1}\cdots (E_{-\varepsilon
_{p},\bar{\varepsilon}_{p}})_{p}\left\vert vac\right\rangle \prod_{j=1}^{p}(-%
\bar{\varepsilon}_{j})  \label{corr}
\end{equation}%
with $\varepsilon _{j},\bar{\varepsilon}_{j}=\pm 1$\ as the vector
components of a function $h_{p}=h_{p}(\lambda _{1},...,\lambda _{p})$ whose
values lie in the tensor space $V^{\otimes 2p}$,$\;V=\mathbb{C}v_{+}\oplus 
\mathbb{C}v_{-}$. Here $\left\vert vac\right\rangle $ denotes the
groundstate of the six-vertex model on an infinite lattice with
inhomogeneity parameters $\{\lambda _{j}\}$. In the homogeneous limit $%
\lambda _{j}\rightarrow 0$ this becomes the groundstate of the Hamiltonian (%
\ref{H})\footnote{%
We follow here the conventions in \cite{BJMST} and consider in the massive
regime, where the the groundstate is twofold-degenerate, only matrix
elements between the same vectors. Notice further that unlike in the case of
the XXX chain it is not known at the moment how to take the homogeneous
directly in the ansatz of \cite{BJMST}.}. The vector valued functions $h_{p}$
are solutions of a set of reduced qKZ equations and according to \cite{BJMST}
can be expressed as%
\begin{equation}
h_{p}(\lambda _{1},...,\lambda _{p})=\frac{e^{\hat{\Omega}_{p}(\lambda
_{1},...,\lambda _{p})}}{2^{p}}\prod_{\ell =1}^{p}\mathfrak{s}_{\ell ,\bar{%
\ell}},\quad \mathfrak{s}=v_{+}\otimes v_{-}-v_{-}\otimes v_{+},\;\bar{\ell}%
=2p+1-\ell 
\end{equation}%
where $\hat{\Omega}=\sum_{i<j}\hat{\Omega}^{(i,j)}$ is a sum of operators of
the form%
\begin{equation}
\hat{\Omega}^{(i,j)}(\lambda _{1},...,\lambda _{p})=\tilde{\omega}(\lambda
_{ij})\tilde{W}^{(i,j)}(q^{\lambda _{1}},...,q^{\lambda _{p}})+\omega
(\lambda _{ij})W^{(i,j)}(q^{\lambda _{1}},...,q^{\lambda _{p}})
\label{omega}
\end{equation}%
with $\lambda _{ij}=\lambda _{i}-\lambda _{j}$ and $\omega ,\tilde{\omega}$
being certain scalar functions. The operators $\tilde{W},W$ depending
rationally on $\{q^{\lambda _{i}}\}$ are defined through the aforementioned
special trace of a monodromy matrix. For instance, setting $(i,j)=(1,2)$ and 
$\lambda _{0}=(\lambda _{1}+\lambda _{2})/2$ one considers the map $%
V^{\otimes 2(p-2)}\rightarrow V^{\otimes 2p}$ 
\begin{equation}
_{p}X_{p-2}(\lambda _{1},...,\lambda _{p})=\frac{\limfunc{Tr}_{\lambda
_{12}}L_{\bar{2}}(\lambda _{02}-1)\cdots L_{\bar{p}}(\lambda _{0\bar{p}%
}-1)L_{p}(\lambda _{0p})\cdots L_{2}(\lambda _{02})}{[\lambda
_{12}]_{q}\prod_{j=3}^{p}[\lambda _{1j}]_{q}[\lambda _{2j}]_{q}}~\mathfrak{s}%
_{1,\bar{2}}\mathfrak{s}_{\bar{1},2}  \label{Xn}
\end{equation}%
which decomposes as%
\begin{equation}
_{p}X_{p-2}(\lambda _{1},...,\lambda _{p})=-\lambda _{12}\cdot _{p}\!\tilde{G%
}_{p-2}(q^{\lambda _{1}},...,q^{\lambda _{p}})+_{p}\!G_{p-2}(q^{\lambda
_{1}},...,q^{\lambda _{p}})\ .  \label{Xdec}
\end{equation}%
The two terms appearing in the above sum determine the two operators on the
right hand side in (\ref{omega}). For further details and the precise
relation between $G,\tilde{G}$ and $W,\tilde{W}$ we refer the reader to \cite%
{BJMST}. For our purposes it will be sufficient to focus only on the
following object in $\limfunc{End}V^{\otimes 2p-2}$ 
\begin{equation}
\boldsymbol{t}:=\limfunc{Tr}\nolimits_{\lambda _{12}}L_{\bar{2}}(\lambda
-\lambda _{2}-1)\cdots L_{\bar{p}}(\lambda -\lambda _{\bar{p}%
}-1)L_{p}(\lambda -\lambda _{p})\cdots L_{2}(\lambda -\lambda _{2})\ .
\label{goal}
\end{equation}%
The operator $L_{j}\in U_{q}(sl_{2})\otimes \limfunc{End}V_{j}$ is a quantum
group intertwiner and 
\begin{equation}
\limfunc{Tr}\nolimits_{\lambda }\equiv \limfunc{Tr}\nolimits_{\lambda ,\zeta
=q^{\lambda }}
\end{equation}%
denotes a linear function%
\begin{equation}
\limfunc{Tr}\nolimits_{\lambda ,\zeta }:U_{q}(sl_{2})\otimes \mathbb{C}%
[\zeta ,\zeta ^{-1}]\rightarrow \lambda \mathbb{C}[\zeta ,\zeta ^{-1}]\oplus 
\mathbb{C}[\zeta ,\zeta ^{-1}]  \label{tracefunc}
\end{equation}%
which for $\lambda =n+1$ yields the conventional trace over the quantum
group representation $\pi ^{(n)}$ of spin $n/2$,%
\begin{equation}
\limfunc{Tr}\nolimits_{\lambda =n+1}(x)=\limfunc{Tr}_{\pi ^{(n)}}x,\quad
\forall x\in U_{q}(sl_{2})\ .  \label{defprop}
\end{equation}%
For our definition of $\pi ^{(n)}$ see (\ref{pin}) in the text below. In
particular, one has for the special elements $x=1,q^{h}$ and the Casimir
operator%
\begin{equation}
\mathfrak{C}=\frac{q^{h-1}+q^{1-h}}{(q-q^{-1})^{2}}+e~f  \label{Cas}
\end{equation}%
the following identities%
\begin{equation}
\limfunc{Tr}\nolimits_{\lambda ,\zeta }~1=\lambda ,\qquad \limfunc{Tr}%
\nolimits_{\lambda ,\zeta }q^{mh}=\frac{\zeta ^{m}-\zeta ^{-m}}{q^{m}-q^{-m}}%
,\qquad \limfunc{Tr}\nolimits_{\lambda ,\zeta }\mathfrak{C}=\frac{\zeta
+\zeta ^{-1}}{(q-q^{-1})^{2}}\ .  \label{special}
\end{equation}%
We will now explain the main point of this article, namely, that there is an
alternative expression for (\ref{goal}) which does not use the introduction
of the abstract trace function (\ref{tracefunc}) but relies on the two
linearly-independent solutions to Baxter's $TQ$ equation for the
inhomogeneous six-vertex model with twisted boundary conditions.

\subsection{Baxter's $TQ$ equation and the six-vertex fusion hierarchy}

Consider the inhomogeneous six-vertex model on a finite lattice with length $%
M=2p-2\in 2\mathbb{N}$, compare with (\ref{goal}). This model can be solved
by finding solutions to Baxter's $TQ$ equation \cite{Bx71,Bx72,Bx73,BxBook},%
\begin{equation}
t(\lambda )Q(\lambda )=Q(\lambda +1)\tprod_{m=1}^{M}[\lambda -\lambda
_{m}]_{q}+Q(\lambda -1)\tprod_{m=1}^{M}[\lambda -\lambda _{m}+1]_{q}\ .
\label{TQ}
\end{equation}%
Here $t$ denotes the six-vertex transfer matrix and $Q$ is an auxiliary
matrix. In terms of eigenvalues (\ref{TQ}) is a difference equation of
second order and hence will in general allow for two linearly independent
solutions, say $Q^{\pm }$ \cite{KWLZ,BLZ,PrSt}. However, not both of them
always possess the same analyticity properties. For instance, for generic $q$
and when $M$ is even, as it is the case here, there is only one solution
which can be expressed as a product of the form 
\begin{equation}
Q^{+}(\lambda )=\tprod_{m=1}^{n_{+}}[\lambda -\xi
_{m}^{+}]_{q}=\tprod_{m=1}^{n_{+}}\frac{\sin \pi \nu \lbrack \lambda -\xi
_{m}^{+}]}{\sin \pi \nu },\qquad \xi _{m}^{+}\equiv \text{ Bethe roots,}
\end{equation}%
the other, $Q^{-}$, contains terms linear in $\lambda $ \cite{PrSt}. The
situation is different for $M$ odd where both solutions take the form of a
product of sine functions. See also the discussion in \cite{KQ4} and \cite%
{FM1,FM2,FM3} for the case when $q$ is a root of unity and the second
solution $Q^{-}$ factorizes into $Q^{+}$ and a "complete string" polynomial
due to a loop algebra symmetry of the model \cite{DFM}. Here we will remove
the associated degeneracies in the spectrum.\smallskip

In this article we give an explicit construction of the $Q$-operators behind
the two solutions $Q^{\pm }$ for even $M$ and relate via a limiting
procedure the linear terms in the second solution to the decomposition in (%
\ref{Xdec}). For the construction of $Q$-operators one in general assumes
analyticity with respect to the variable $z=q^{2\lambda }$ by requiring that
the $Q$-operators commute for arbitrary values of the spectral parameter, $%
[Q(z_{1}),Q(z_{2})]=0$ \cite{BxBook}. This usually prevents the appearance
of linear terms in $\lambda $.

The key to obtain these terms is the introduction of quasi-periodic boundary
conditions on the finite lattice taking the limit to periodic boundary
conditions at the very end of the construction. The twisted boundary
conditions depend on a generic twist parameter $\alpha $ where the value $%
\alpha =0$ corresponds to ordinary periodic boundary conditions. For $\alpha
\neq 0$ two linearly independent solutions to (\ref{TQ}) without linear
terms can be constructed and one finds a simplified expression for the
six-vertex fusion hierarchy: denote by $t_{\alpha }^{(n)}$ the transfer
matrix with spin $n/2$ in the auxiliary space then one has%
\begin{equation}
t_{\alpha }^{(n-1)}(\lambda )=\frac{q^{-n\alpha }Q_{\alpha }^{+}(\lambda +%
\frac{n}{2})Q_{\alpha }^{-}(\lambda -\frac{n}{2})-q^{n\alpha }Q_{\alpha
}^{+}(\lambda -\frac{n}{2})Q_{\alpha }^{-}(\lambda +\frac{n}{2})}{%
q^{-S^{z}-\alpha }-q^{\alpha +S^{z}}}\ .  \label{fusion}
\end{equation}%
Here $S^{z}$ denotes the total spin operator acting on the chain. By
analytic continuation of this formula with respect to the spin variable $n/2$
of the transfer matrix one obtains in the limit $\alpha \rightarrow 0$ an
alternative expression for (\ref{goal}),%
\begin{equation}
\boldsymbol{t}(\lambda ,\zeta )=\lim_{\alpha \rightarrow 0}\frac{\zeta
^{-\alpha }Q_{\alpha }^{+}(q^{\lambda }\zeta ^{\frac{1}{2}})Q_{\alpha
}^{-}(q^{\lambda }\zeta ^{-\frac{1}{2}})-\zeta ^{\alpha }Q_{\alpha
}^{+}(q^{\lambda }\zeta ^{-\frac{1}{2}})Q_{\alpha }^{-}(q^{\lambda }\zeta ^{%
\frac{1}{2}})}{q^{-S^{z}-\alpha }-q^{\alpha +S^{z}}}\ ,  \label{result}
\end{equation}%
which reproduces all the desired properties (\ref{special}) of the trace
function, in particular the appearance of linear terms in $\lambda $,%
\begin{equation}
q^{M\lambda }\boldsymbol{t}(\lambda ,\zeta =q^{\lambda })=\lambda \cdot 
\boldsymbol{\tilde{g}}(q^{2\lambda })+\boldsymbol{g}(q^{2\lambda })\ .
\label{result2a}
\end{equation}%
In (\ref{result}) we have identified on the right hand side of the equation $%
Q_{\alpha }^{\pm }(\lambda )\equiv Q_{\alpha }^{\pm }(q^{\lambda })$ and $%
\boldsymbol{\tilde{g}}(x),\boldsymbol{g}(x)$ in (\ref{result2a}) are
operator valued polynomials whose degrees in each fixed spin sector $S^{z}$
are given by%
\begin{equation}
\deg \boldsymbol{\tilde{g}}|_{S^{z}}=M-|S^{z}|\qquad \text{and}\qquad \deg 
\boldsymbol{g}|_{S^{z}}\leq M\ ,  \label{result2b}
\end{equation}%
where the upper bound for $\deg \boldsymbol{g}|_{S^{z}}$ is assumed when $%
S^{z}\neq 0$ but it is strictly smaller than $M$ when $S^{z}=0$. This will
be derived in Section 4. Before that we present in Section 3 our concrete
construction of the $Q$-operators, which will relate the appearance of the
extra variable $\zeta $ to a special restriction of a Verma module of the
upper Borel subalgebra $U_{q}(b_{+})\subset U_{q}(\widehat{sl}_{2})$%
.\smallskip

It should be emphasized that our construction of $Q$-operators for the
quasi-periodic chain has applications beyond the focus of this article and
it would be interesting to see whether similar formulae also occur in the
treatment of the correlation functions for the finite XXZ chain \cite%
{KMT,KMST,GKS}. These approaches rely on the Bethe ansatz and we stress that
the Bethe ansatz equations are also contained in (\ref{fusion}) when setting 
$n=1$ and shifting $\lambda \rightarrow \lambda +1/2$,%
\begin{equation}
\prod_{m=1}^{M}[\lambda -\lambda _{m}+1]_{q}=\frac{q^{-\alpha }Q_{\alpha
}^{+}(\lambda +1)Q_{\alpha }^{-}(\lambda )-q^{\alpha }Q_{\alpha
}^{+}(\lambda )Q_{\alpha }^{-}(\lambda +1)}{q^{-S^{z}-\alpha }-q^{\alpha
+S^{z}}}\ .  \label{w}
\end{equation}%
In the context of the Liouville model the analogue of this identity has been
referred to as quantum Wronskian \cite{BLZ}. Evaluating it at $\lambda =\xi
_{m}^{\pm },\xi _{m}^{\pm }-1$ gives the Bethe ansatz equations above and
below the equator \cite{KWLZ,PrSt}. However, the above equation (\ref{w}) is
of a simpler form than the Bethe ansatz equations and it is desirable to
understand its consequences in the thermodynamic limit.

An analogous construction of $Q$-operators for the eight-vertex or $XYZ$
model is currently under investigation \cite{inprep}.

\section{The six-vertex fusion hierarchy}

The inhomogeneous six-vertex model on a finite lattice is associated with
the quantum loop algebra $U_{q}(\widetilde{sl}_{2})$ in terms of which the
transfer matrix and fusion hierarchy can be defined. Consider the spin $n/2$
representation $\pi ^{(n)}$ of the finite quantum subgroup $%
U_{q}(sl_{2})\subset U_{q}(\widetilde{sl}_{2})$, i.e. 
\begin{eqnarray}
\pi ^{(n)}(e)\left\vert k\right\rangle &=&[n-k+1]_{q}[k]_{q}\left\vert
k-1\right\rangle ,\quad  \label{pin} \\
\pi ^{(n)}(f)\left\vert k\right\rangle &=&\left\vert k+1\right\rangle ,\quad
\notag \\
\pi ^{(n)}(q^{\frac{h}{2}})\left\vert k\right\rangle &=&q^{\frac{n}{2}%
-k}\left\vert k\right\rangle  \notag
\end{eqnarray}%
with $k=0,1,...,n$. The transfer matrices of the model are built up from the
intertwiners of the respective evaluation modules. Define 
\begin{equation}
L=\left( 
\begin{array}{cc}
z~q^{\frac{h}{2}+1}-q^{-\frac{h}{2}} & z~(q-q^{-1})q^{\frac{h}{2}+1}f \\ 
(q-q^{-1})eq^{-\frac{h}{2}} & z~q^{-\frac{h}{2}+1}-q^{\frac{h}{2}}%
\end{array}%
\right) \in U_{q}(sl_{2})\otimes \limfunc{End}V  \label{L0}
\end{equation}%
where $V$ is identified as the two-dimensional representation space of $\pi
^{(1)}$ and $z=q^{2\lambda }$ as the spectral parameter. Then the
inhomogeneous transfer matrix of spin $n/2$ is introduced by setting%
\footnote{%
Notice that we have changed our conventions in comparison with \cite%
{KQ3,KQ4,KQ5} and have made the following change in notation, $%
T^{(n)}(z)\rightarrow T^{(n-1)}(zq^{n})$.} 
\begin{equation}
T_{\alpha }^{(n)}(z)=\limfunc{Tr}_{\pi ^{(n)}}q^{\alpha h\otimes
1}L_{M}(z\zeta _{M}^{2})\cdots L_{1}(z\zeta _{1}^{2})\in \limfunc{End}%
V^{\otimes M}\ .  \label{Tn}
\end{equation}%
For the moment the length of the spin-chain $M$ can be any positive integer
and $\alpha $ is the twist angle. The set $\{\zeta _{m}=q^{-\lambda
_{m}}\}_{m=1}^{M}$ are some unspecified "generic" inhomogeneity parameters.
The above transfer matrices constitute the six-vertex fusion hierarchy
satisfying the functional equation \cite{KR} 
\begin{equation}
T_{\alpha }^{(n)}(zq^{n+1})T_{\alpha }^{(1)}(z)=T_{\alpha
}^{(n+1)}(zq^{n})\prod_{m=1}^{M}\left( zq^{2}\zeta _{m}^{2}-1\right)
+T_{\alpha }^{(n-1)}(zq^{n+2})\prod_{m=1}^{M}(z\zeta _{m}^{2}-1)\;.
\label{fus}
\end{equation}%
In the homogeneous limit, $\zeta _{m}\rightarrow 1$, we obtain from the
transfer matrix $T$ as logarithmic derivative the Hamiltonian for the finite 
$XXZ$ chain, 
\begin{eqnarray}
H_{\text{XXZ}} &=&-\left( q-q^{-1}\right) ~\left. z\frac{d}{dz}\ln \frac{%
T_{\alpha }(z)}{(zq^{2}-1)^{M}}\right\vert _{z=1}  \notag \\
&=&-\frac{1}{2}\sum_{m=1}^{M}\left\{ \sigma _{m}^{x}\sigma _{m+1}^{x}+\sigma
_{m}^{y}\sigma _{m+1}^{y}+\frac{q+q^{-1}}{2}\,(\sigma _{m}^{z}\sigma
_{m+1}^{z}-1)\right\} \;.
\end{eqnarray}%
Here the twisted boundary conditions manifest itself in the identification%
\begin{equation*}
\sigma _{M+1}^{\pm }=q^{\pm 2\alpha }\sigma _{1}^{\pm }\ .
\end{equation*}%
The well-known symmetries of the model are expressed in terms of the
following commutators 
\begin{equation}
\lbrack T_{\alpha }^{(m)}(z),T_{\alpha }^{(n)}(w)]=[T_{\alpha
}^{(n)}(z),S^{z}]=[T_{\alpha }^{(n)}(z),\mathfrak{S}]=0\;,  \label{symm}
\end{equation}%
and%
\begin{equation}
\mathfrak{R}T_{\alpha }^{(n)}(z)=T_{-\alpha }^{(n)}(z)\mathfrak{R}
\end{equation}%
where the respective operators are defined as 
\begin{equation}
S^{z}=\frac{1}{2}\sum_{m=1}^{M}\sigma _{m}^{z},\text{\quad }\mathfrak{R}%
=\sigma ^{x}\otimes \cdots \otimes \sigma ^{x}=\prod\limits_{m=1}^{M}\sigma
_{m}^{x},\text{\quad }\mathfrak{S}=\sigma ^{z}\otimes \cdots \otimes \sigma
^{z}=\prod\limits_{m=1}^{M}\sigma _{m}^{z}\;.  \label{SzRS}
\end{equation}%
These symmetries hold for spin-chains of even as well as odd length. For
later purposes we also compute the value of the transfer matrices at the
origin,%
\begin{equation}
T_{\alpha }^{(n-1)}(0)=(-)^{M}\limfunc{Tr}_{\pi ^{(n-1)}}q^{(\alpha
-S^{z})h}=(-)^{M}\frac{q^{n(S^{z}-\alpha )}-q^{-n(S^{z}-\alpha )}}{%
q^{S^{z}-\alpha }-q^{-S^{z}+\alpha }}\ .
\end{equation}

\section{The auxiliary matrices}

Our construction follows the main steps of the previous works \cite{RW02}
and \cite{KQ,KQ2,KQ3,KQ4,KQ5} with some minor modifications. We therefore
only state the main results and briefly comment on the steps involved to
accommodate twisted boundary conditions. The reader is referred to the
aforementioned literature for the technical details.

\subsection{Definition}

For the definition of the auxiliary matrix we employ the following
representations $\pi ^{\pm }=\pi ^{\pm }(z;r,s)$ of the upper Borel
subalgebra $U_{q}(b_{+})\subset U_{q}(\widetilde{sl}_{2})$. Let $k\in 
\mathbb{N}_{\geq 0}$ then%
\begin{eqnarray}
\pi ^{+}(e_{0})\left\vert k\right\rangle &=&z\left\vert k+1\right\rangle
,\quad \pi ^{+}(q^{\frac{h_{1}}{2}})\left\vert k\right\rangle =\pi ^{+}(q^{-%
\frac{h_{0}}{2}})\left\vert k\right\rangle =rq^{-2k-1}\left\vert
k\right\rangle ,  \notag \\
\pi ^{+}(e_{1})\left\vert k\right\rangle &=&\frac{s+1-q^{2k}-sq^{-2k}}{%
(q-q^{-1})^{2}}\;\left\vert k-1\right\rangle ,\quad \pi
^{+}(e_{1})\left\vert 0\right\rangle =0,\quad r,s,z\in \mathbb{C}
\label{piplus}
\end{eqnarray}%
and set 
\begin{equation}
\pi ^{-}:=\pi ^{+}\circ \omega \quad \text{with\quad }\{e_{1},e_{0},q^{\frac{%
h_{1}}{2}},q^{\frac{h_{0}}{2}}\}\overset{\omega }{\rightarrow }%
\{e_{0},e_{1},q^{\frac{h_{0}}{2}},q^{\frac{h_{1}}{2}}\}\;.
\end{equation}%
The representation $\pi ^{-}$ is a particular restriction of the
representation introduced in \cite{RW02}. When $q$ is a primitive root of
unity of order $N$ we set $N^{\prime }=N$ if the order is odd and $N^{\prime
}=N/2$ if it is even. Then the above infinite-dimensional representation
becomes reducible and can be truncated \cite{KQ5},%
\begin{equation}
\pi ^{+}(e_{0})\left\vert N^{\prime }-1\right\rangle =0\ .  \label{trunc}
\end{equation}%
For roots of unity this truncation will always be implicitly understood. In
order to unburden the formulae we will often drop the explicit dependence on
the parameters $\{z,r,s\}$ and the representation $\pi ^{+}$. Set 
\begin{equation}
\mathfrak{L}=\left( 
\begin{array}{cc}
z~\frac{s}{r}q^{\frac{h_{1}}{2}+1}-q^{-\frac{h_{1}}{2}} & (q-q^{-1})e_{0}q^{-%
\frac{h_{0}}{2}} \\ 
(q-q^{-1})e_{1}q^{-\frac{h_{1}}{2}} & zr~q^{-\frac{h_{1}}{2}+1}-q^{\frac{%
h_{1}}{2}}%
\end{array}%
\right) \in U_{q}(b_{+})\otimes \limfunc{End}V,  \label{Ldec}
\end{equation}%
then $\mathfrak{L}_{\pi ^{+}}=(\pi ^{+}\otimes 1)\mathfrak{L}$ is the
intertwiner of the tensor product $\pi ^{+}\otimes \pi ^{(1)}$. Notice that
the intertwiner for the representation $\pi ^{-}$ is obtained via spin
reversal, i.e. 
\begin{equation}
\mathfrak{L}_{\pi ^{-}}=(1\otimes \sigma ^{x})\mathfrak{L}_{\pi
^{+}}(1\otimes \sigma ^{x})\;.
\end{equation}%
Define the auxiliary matrix in terms of the intertwiner $\mathfrak{L}$ as
the trace of the following operator product, 
\begin{equation}
Q_{\alpha }(z;r,s)=\limfunc{Tr}_{\pi ^{+}}q^{\alpha h_{1}\otimes 1}\mathfrak{%
L}_{M}(z\zeta _{M}^{2};r,s)\cdots \mathfrak{L}_{1}(z\zeta _{1}^{2};r,s)\ .
\label{Qmu}
\end{equation}%
For later purposes we also define the special limits%
\begin{equation}
Q_{\alpha }^{+}(z)=\lim_{s\rightarrow 0}Q_{\alpha }(0;1,s)^{-1}Q_{\alpha
}(z;1,s)\quad \text{and\quad }Q_{\alpha }^{-}=\mathfrak{R}Q_{-\alpha }^{+}%
\mathfrak{R\ }\text{.}  \label{Qpm}
\end{equation}%
By definition of the auxiliary matrix the operators $Q_{\alpha }^{\pm }$ are
well defined and we have for roots of unity%
\begin{equation}
q^{N}=1:\qquad Q_{\alpha }(0;r,s)=(-)^{M}r^{\alpha -S^{z}}\frac{%
1-q^{2N^{\prime }(S^{z}-\alpha )}}{q^{\alpha -S^{z}}-q^{S^{z}-\alpha }}~,
\end{equation}%
while for 
\begin{equation}
\text{generic }q:\qquad Q_{\alpha }(0;r,s)=\frac{(-)^{M}r^{\alpha -S^{z}}}{%
q^{\alpha -S^{z}}-q^{S^{z}-\alpha }}\ \text{\quad with\quad }|q^{-\alpha \pm 
\frac{M}{2}}|<1\text{ for }|q|^{\pm 1}>1\ .
\end{equation}%
After a suitable renormalization (see Section 4.1) the eigenvalues of the
operators (\ref{Qpm}) yield the two linearly independent solutions to
Baxter's TQ equation mentioned in the introduction.

\subsection{Spin conservation and reversal}

The matrix (\ref{Qmu}) commutes by construction with the fusion hierarchy
and preserves two of the symmetries (\ref{symm}) \cite{KQ,KQ2}, 
\begin{equation}
\lbrack Q_{\alpha }(z;r,s),S^{z}]=[Q_{\alpha }(z;r,s),\mathfrak{S}]=0\ .
\label{QSS}
\end{equation}%
Because of spin conservation the dependence of the auxiliary matrix $Q$ on
the parameter $r$ can be easily extracted,%
\begin{equation}
Q_{\alpha }(z;r,s)=r^{\alpha -S^{z}}Q_{\alpha }(z;1,s)\equiv r^{\alpha
-S^{z}}Q_{\alpha }(z;s)
\end{equation}%
Without loss of generality we can therefore set $r=1$ in order to discuss
the behaviour of the auxiliary matrix under spin reversal. Employing once
more spin conservation we easily find from%
\begin{eqnarray*}
\left\langle v_{+}\right\vert \mathfrak{L}\left\vert v_{+}\right\rangle
&=&zs~q^{1+\frac{h_{1}}{2}}-q^{-\frac{h_{1}}{2}%
}=(-zsq)(z^{-1}q^{-2}s^{-1}~q^{1-\frac{h_{1}}{2}}-q^{\frac{h_{1}}{2}}) \\
\left\langle v_{-}\right\vert \mathfrak{L}\left\vert v_{-}\right\rangle
&=&z~q^{1-\frac{h_{1}}{2}}-q^{\frac{h_{1}}{2}%
}=(-zq)(z^{-1}q^{-2}s^{-1}~sq^{1+\frac{h_{1}}{2}}-q^{-\frac{h_{1}}{2}})
\end{eqnarray*}%
the identities 
\begin{equation}
\mathfrak{R}Q_{\alpha }(z;s,\{\zeta _{m}\})\mathfrak{R}=(-z)^{M}q^{M}s^{%
\frac{M}{2}-S^{z}}Q_{\alpha }(z^{-1}q^{-2}s^{-1};s,\{\zeta
_{m}^{-1}\})^{t}\tprod_{m}\zeta _{m}^{2}  \label{RQR1}
\end{equation}%
and%
\begin{equation}
\mathfrak{R}Q_{\alpha }(z,q;s)\mathfrak{R}=Q_{-\alpha
}(zq^{2}s,q^{-1};s^{-1})^{t}\ .  \label{RQR2}
\end{equation}

\subsection{Commutation of the auxiliary matrices}

For generic $q$ we can employ the concept of the universal $R$-matrix and
the fact that $\pi ^{+}(z;r,s)$ is the restriction of a well-known
evaluation Verma module of $U_{q}(\widehat{sl}_{2})$ to conclude that the
intertwiner, say $S$, of the tensor product%
\begin{equation*}
\pi ^{+}(z;r,s)\otimes \pi ^{+}(w;r^{\prime },s^{\prime })
\end{equation*}%
exists. Consequently, the Yang-Baxter relation $S_{12}\mathfrak{L}_{13}%
\mathfrak{L}_{23}^{\prime }=\mathfrak{L}_{23}^{\prime }\mathfrak{L}%
_{13}S_{12}$ holds and the auxiliary matrices commute among each other, 
\begin{equation}
\lbrack Q_{\alpha }(z;r,s),Q_{\alpha }(w;r^{\prime },s^{\prime })]=0\ .
\label{QQ}
\end{equation}%
Here we have implicitly used that 
\begin{equation}
\lbrack \mathfrak{L},\pi ^{+}(h_{1})\otimes \sigma ^{z}]=[S,\pi
^{+}(h_{1})\otimes \pi ^{+}(h_{1})^{\prime }]=0
\end{equation}%
in order to ensure compatibility with the quasi-periodic boundary
conditions. The above commutation relations are a direct consequence of the
intertwining property of $\mathfrak{L}$ and $S$.

When $q$ is a root of unity we cannot use the existence of the universal $R$%
-matrix and have to construct the intertwiners $S$ explicitly. This has been
done for $N=3,4,6$ in \cite{KQ2,KQ5}. Numerical checks have been carried out
for $N=5,7,8$ and the commutation relation was found to hold. \emph{We
therefore assume as a working hypothesis that the intertwiner }$S$\emph{\
also exists in the root of unity case for general }$N$\emph{.}

From (\ref{QQ}) it follows that the eigenvalues of the auxiliary matrix are
polynomial in the parameters $z,s$ and that the eigenvectors of the $Q$%
-operator do not depend on them.

\subsection{The $TQ$ equation}

One of the essential properties of the auxiliary matrix $Q_{\alpha }$ is
that it satisfies the functional equation%
\begin{eqnarray}
T_{\alpha }(z)Q_{\alpha }(z;r,s) &=&  \notag \\
&&\hspace{-2.5cm}Q_{\alpha }(zq^{2};rq^{-1},sq^{-2})\prod_{m}(z\zeta
_{m}^{2}-1)+Q_{\alpha }(zq^{-2};rq,sq^{2})\prod_{m}(z\zeta _{m}^{2}q^{2}-1)
\end{eqnarray}%
which one derives from the following non-split exact sequence of
representations \cite{RW02,KQ} 
\begin{equation}
0\rightarrow \pi ^{+}(zq^{2};rq^{-1},sq^{-2})\hookrightarrow \pi
^{+}(z;r,s)\otimes \pi _{z}^{(1)}\rightarrow \pi
^{+}(zq^{-2},rq,sq^{2})\rightarrow 0\ .  \label{seq1}
\end{equation}%
The proof can be found in \cite{RW02, KQ}, here we have only to incorporate
twisted boundary conditions. To this end note for instance that the
inclusion $\imath :\pi ^{+}(zq^{2};rq^{-1},sq^{-2})\hookrightarrow \pi
^{+}(z;r,s)\otimes \pi _{z}^{(1)}$ in (\ref{seq}) is given in the following
form 
\begin{equation*}
\left\vert k\right\rangle \hookrightarrow \left\vert k+1\right\rangle
\otimes v_{+}+c_{k}\left\vert k\right\rangle \otimes v_{-}
\end{equation*}%
where $c_{k}$ is some coefficient whose explicit form is not relevant here.
Then one easily verifies that 
\begin{equation*}
\imath \circ \pi ^{+}(q^{\alpha h_{1}};z,rq^{-1},sq^{-2})=\pi ^{+}(q^{\alpha
h_{1}};z,r,s)\circ \imath \ .
\end{equation*}%
Similarly one shows that the projection $p:\pi ^{+}(z;r,s)\otimes \pi
_{z}^{(1)}\rightarrow \pi ^{+}(zq^{-2},rq,sq^{2}),$%
\begin{equation*}
p:~\left\vert k\right\rangle \mapsto c_{k}^{\prime }~\left\vert
k\right\rangle \otimes v_{+}
\end{equation*}
in (\ref{seq}) commutes with the twist operator as well,%
\begin{equation*}
p\circ \pi ^{+}(q^{\alpha h_{1}};z,r,s)=\pi ^{+}(q^{\alpha
h_{1}};z,rq,sq^{2})\circ p\ .
\end{equation*}

Setting $r=1$ and taking the limit $s\rightarrow 0$ we find%
\begin{equation*}
T_{\alpha }(z)Q_{\alpha }^{\pm }(z)=q^{\pm (S^{z}-\alpha )}Q_{\alpha }^{\pm
}(zq^{2})\prod_{m}(z\zeta _{m}^{2}-1)+q^{\pm (\alpha -S^{z})}Q_{\alpha
}^{\pm }(zq^{-2})\prod_{m}(z\zeta _{m}^{2}q^{2}-1)\ .
\end{equation*}%
As we will see below, this relation is identical to Baxter's famous $TQ$
equation mentioned in the introduction after an appropriate rescaling of the
auxiliary matrix; see Section 4.1. The eigenvalues of the operators $%
Q_{\alpha }^{\pm }$ defined in (\ref{Qpm}) then coincide with the two
linearly independent solutions of (\ref{TQ}) discussed in the introduction.

As an immediate consequence of the $TQ$ equation and the fusion relation we
have in terms of eigenvalues the identity%
\begin{equation}
T_{\alpha }^{(n-1)}(z)=q^{-(n+1)(\alpha -S^{z})}Q_{\alpha
}^{+}(zq^{-n})Q_{\alpha }^{+}(zq^{n})\sum_{\ell =1}^{n}\frac{q^{2\ell
(\alpha -S^{z})}\prod_{m}(z\zeta _{m}^{2}q^{2\ell -n}-1)}{Q_{\alpha
}^{+}(zq^{2\ell -n})Q_{\alpha }^{+}(zq^{2\ell -n-2})}\ .
\end{equation}%
This formula is easily proved by induction.

\subsection{Functional equation at roots of unity}

When $q$ is a root of unity the vital information on the spectrum of the
auxiliary matrix is encoded in the following functional equation which is a
straightforward generalization of a previous result \cite{KQ4} to twisted
boundary conditions\footnote{%
In order to facilitate the comparison with \cite{KQ4,KQ5} note that we have
made the following changes. In order to match the result in \cite{KQ4} set $%
s=\mu ^{-2}$ and $r=\mu ^{-1}$. We have also redefined the fusion hierarchy, 
$T^{(n)}(z)\rightarrow T^{(n-1)}(zq^{n})$, and shifted the parameter $r$ in 
\cite{KQ5} by $r\rightarrow rq^{-1}$.},%
\begin{align}
Q_{\alpha }(zq^{2}/s;s)Q_{\alpha }(z;t)& =  \notag \\
& \hspace{-2cm}q^{S^{z}-\alpha }Q_{\alpha }(zq^{2}/s;stq^{-2})\left[
\prod_{m}(z\zeta _{m}^{2}q^{2}-1)+q^{N^{\prime }(S^{z}-\alpha )}T_{\alpha
}^{(N^{\prime }-2)}(zq^{N^{\prime }+1})\right] \ .  \label{QQQ}
\end{align}%
It implies the following decomposition of the eigenvalues%
\begin{equation}
Q_{\alpha }(z;s)=Q_{\alpha }(0)Q_{\alpha }^{+}(z)Q_{\alpha }^{-}(zs),\quad
Q_{\alpha }^{\pm }(z)=\prod_{i=1}^{n_{\pm }}(1-z~x_{i}^{\pm })
\end{equation}%
where the last factors $Q_{\alpha }^{\pm }$ are polynomials with a total of $%
M=n_{+}+n_{-}$ roots and are related by the following formula%
\begin{equation}
Q_{\alpha }(0)Q_{\alpha }^{-}(z)=q^{(2N^{\prime }+1)(S^{z}-\alpha
)}Q_{\alpha }^{+}(z)\sum_{\ell =1}^{N^{\prime }}\frac{q^{2\ell (\alpha
-S^{z})}\prod_{m}(z\zeta _{m}^{2}q^{2\ell }-1)}{Q_{\alpha }^{+}(zq^{2\ell
})Q_{\alpha }^{+}(zq^{2\ell -2})}\ .  \label{QsumQ}
\end{equation}%
Employing the transformation laws (\ref{RQR1}), (\ref{RQR2}) of the
auxiliary matrix under spin-reversal we deduce that%
\begin{equation}
n_{\pm }=\frac{M}{2}\mp S^{z}\ .
\end{equation}%
From the above identity (\ref{QsumQ}) between the eigenvalues of $Q_{\alpha
}^{\pm }$ we now derive the following expression for the fusion hierarchy%
\begin{multline*}
q^{-n(\alpha -S^{z})}Q_{\alpha }(0)Q_{\alpha }^{+}(zq^{2n})Q_{\alpha
}^{-}(z)-q^{n(\alpha -S^{z})}Q_{\alpha }(0)Q_{\alpha }^{+}(z)Q_{\alpha
}^{-}(zq^{2n})= \\
q^{(2N^{\prime }+1+n)(S^{z}-\alpha )}Q_{\alpha }^{+}(zq^{2n})Q_{\alpha
}^{+}(z)\tsum_{\ell =1}^{N^{\prime }}\tfrac{q^{2\ell (\alpha -S^{z})}\prod
(z\zeta _{m}^{2}q^{2\ell }-1)}{Q_{\alpha }^{+}(zq^{2\ell })Q_{\alpha
}^{+}(zq^{2\ell -2})} \\
-q^{(2N^{\prime }+1+n)(S^{z}-\alpha )}Q_{\alpha }^{+}(z)Q_{\alpha
}^{+}(zq^{2n})\left\{ \tsum_{\ell =n+1}^{N^{\prime }}...+q^{2N^{\prime
}(\alpha -S^{z})}\sum_{\ell =1}^{n}...\right\} = \\
(q^{2N^{\prime }(S^{z}-\alpha )}-1)T_{\alpha }^{(n-1)}(zq^{n})
\end{multline*}%
After inserting the value for $Q_{\alpha }(0)$ we obtain%
\begin{equation}
T_{\alpha }^{(n-1)}(z)=(-)^{M}\frac{q^{n(S^{z}-\alpha )}Q_{\alpha
}^{+}(zq^{n})Q_{\alpha }^{-}(zq^{-n})-q^{n(\alpha -S^{z})}Q_{\alpha
}^{+}(zq^{-n})Q_{\alpha }^{-}(zq^{n})}{q^{S^{z}-\alpha }-q^{\alpha -S^{z}}}\
.  \label{TnQQ}
\end{equation}%
As long as $\alpha \neq 0\func{mod}N^{\prime },S^{z}$ this expression holds
true for $M,N$ even and odd. In particular we have for $n=1$ the identity%
\begin{equation}
\prod_{m=1}^{M}(1-z\zeta _{m}^{2}q)=\frac{q^{S^{z}-\alpha }Q_{\alpha
}^{+}(zq)Q_{\alpha }^{-}(zq^{-1})-q^{\alpha -S^{z}}Q_{\alpha
}^{+}(zq^{-1})Q_{\alpha }^{-}(zq)}{q^{S^{z}-\alpha }-q^{\alpha -S^{z}}}
\end{equation}%
which has been called "Quantum Wronskian" in the literature \cite{BLZ}.

\subsection{The algebraic Bethe ansatz for generic $q$}

In \cite{KQ3} the spectrum of auxiliary matrices constructed in \cite{RW02}
have been investigated on the basis of the algebraic Bethe ansatz. The
results in \cite{KQ3} apply to the inhomogeneous XXZ-chain with twisted
boundary conditions. Denote the monodromy matrices associated with the
transfer and auxiliary matrix by%
\begin{equation*}
\mathcal{T}=(\pi ^{(1)}\otimes 1)q^{\alpha h\otimes 1}L_{M}\cdots
L_{1}=\left( 
\begin{array}{cc}
\mathcal{A} & \mathcal{B} \\ 
\mathcal{C} & \mathcal{D}%
\end{array}%
\right)
\end{equation*}%
and%
\begin{equation*}
\mathcal{Q}=(\pi ^{+}\otimes 1)q^{\alpha h_{1}\otimes 1}\mathfrak{L}%
_{M}\cdots \mathfrak{L}_{1}=(\mathcal{Q}_{k\ell })_{k,\ell \geq 0},\text{%
\quad }\mathcal{Q}_{k\ell }=\left\langle k\right\vert \mathcal{Q}\left\vert
\ell \right\rangle ,
\end{equation*}%
respectively. If the quantum space $\mathcal{H}$ carries a highest (or
lowest) weight representation of the quantum group with highest weight
vector $\left\vert 0\right\rangle _{\mathcal{H}}$ then the eigenvectors and
eigenvalues of the $Q$-operator can be computed via the algebraic Bethe
ansatz. Setting $r=1$ in (\ref{piplus}) and denoting by $\mathfrak{L}%
_{k,\varepsilon }^{\ell ,\varepsilon ^{\prime }}=\left\langle \ell
,\varepsilon ^{\prime }|\mathfrak{L}|k,\varepsilon \right\rangle $ the
matrix elements of the $\mathfrak{L}$-operator one finds%
\begin{multline*}
Q_{\alpha }(z)\prod_{j=1}^{n_{+}}\mathcal{B}(x_{j}^{+})\left\vert
0\right\rangle _{\mathcal{H}}= \\
\left\{ \sum_{k\geq 0}\,\left\langle 0|\mathcal{Q}_{kk}(z)|0\right\rangle _{%
\mathcal{H}}\prod_{j=1}^{n_{+}}\left( \frac{\mathfrak{L}%
_{k,-}^{k,-}(zx_{j}^{+})}{\mathfrak{L}_{k,+}^{k,+}(zx_{j}^{+})}-\frac{%
\mathfrak{L}_{k,-}^{k+1,+}(zx_{j}^{+})\mathfrak{L}_{k+1,+}^{k,-}(zx_{j}^{+})%
}{\mathfrak{L}_{k+1,+}^{k+1,+}(zx_{j}^{+})\mathfrak{L}%
_{k,+}^{k,+}(zx_{j}^{+})}\right) \right\} \,\prod_{j=1}^{n_{+}}\mathcal{B}%
(x_{j}^{+})\left\vert 0\right\rangle _{\mathcal{H}}= \\
\left\{ \sum_{k\geq 0}\,\left\langle 0|\mathcal{Q}_{kk}(z)|0\right\rangle _{%
\mathcal{H}}\prod_{j=1}^{n_{+}}\frac{q^{-2k-1}(1-zx_{j}^{+})(1-zx_{j}^{+}s)}{%
(1-zx_{j}^{+}sq^{-2k})(1-zx_{j}^{+}sq^{-2k-2})}\right\} \,\prod_{j=1}^{n_{+}}%
\mathcal{B}(x_{j}^{+})\left\vert 0\right\rangle _{\mathcal{H}},
\end{multline*}%
where $x_{j}^{+}$ are the Bethe roots above the equator. This formula has
been proved directly for $n_{+}=1,2,3$ in \cite{KQ3} and further checked for
consistency for arbitrary $n_{+}$. In the present case of the XXZ chain we
have%
\begin{equation}
\left\vert 0\right\rangle _{\mathcal{H}}=v_{+}\otimes \cdots \otimes
v_{+},\quad \mathcal{H}=(\mathbb{C}^{2})^{\otimes M}
\end{equation}%
and%
\begin{equation}
\left\langle 0|\mathcal{Q}_{kk}(z)|0\right\rangle _{\mathcal{H}}=q^{-\alpha
(2k+1)}\prod_{m}q^{k+\frac{1}{2}}(zs\zeta _{m}^{2}q^{-2k}-1)\ .
\end{equation}%
The above expression from the algebraic Bethe ansatz then yields for the
case of generic $q$ the same information on the spectrum which we had
previously derived from the functional equation (\ref{QQQ}) at a root of
unity,%
\begin{multline*}
Q_{\alpha }(z)\prod_{j=1}^{n_{+}}\mathcal{B}(x_{j}^{+})\left\vert
0\right\rangle _{\mathcal{H}}= \\
\left\{ q^{S^{z}-\alpha }Q_{\alpha }^{+}(z)~Q_{\alpha }^{+}(zs)\sum_{k\geq 0}%
\frac{q^{2k(S^{z}-\alpha )}\prod (zs\zeta _{m}^{2}q^{-2k}-1)}{Q_{\alpha
}^{+}(zsq^{-2k})Q_{\alpha }^{+}(zsq^{-2k-2})}\right\} ~\prod_{j=1}^{n_{+}}%
\mathcal{B}(x_{j}^{+})\left\vert 0\right\rangle _{\mathcal{H}},
\end{multline*}%
where we have set as before%
\begin{equation}
Q_{\alpha }^{+}(z)=\prod_{j=1}^{n_{+}}(1-zx_{j}^{+}),\qquad n_{+}=\frac{M}{2}%
-S^{z}\ .
\end{equation}%
Thus, the analogous result concerning the decomposition of the eigenvalues
into $Q_{\alpha }^{\pm }$ applies also here. Notice that for generic $q$ one
has to put further restrictions on the twist parameter $\alpha $ \cite{KQ3},%
\begin{equation*}
|q^{-\alpha }{}^{\pm M/2}|<1\qquad \text{for\qquad }|q|^{\pm 1}\geq 1,
\end{equation*}%
in order to ensure absolute convergence. Analytically continuing from this
region the relation (\ref{TnQQ}) holds also true for generic $q$.

An alternative to the Bethe ansatz is to follow the analogous line of
argument as presented in \cite{BLZ}. Note that when we set $s=q^{2n}$ in (%
\ref{piplus}) it follows that%
\begin{equation}
s=q^{2n}:\qquad e_{1}\left\vert n\right\rangle =0\ .
\end{equation}%
Hence, the infinite-dimensional module $\pi ^{+}$ splits into a finite $n$%
-dimensional part $W_{<n}$ given by the linear span of the basis vectors $%
\{\left\vert k\right\rangle \}_{k=0}^{n-1}$ and an infinite-dimensional
space $W_{\geq n}$ which is the linear span of the vectors $\{\left\vert
k\right\rangle \}_{k=n}^{\infty }$. Note that $W_{\geq n}$ is left invariant
under the action of $U_{q}(b_{+})$. Restricting the $\mathfrak{L}$-operator
onto these two spaces, one finds%
\begin{equation}
\left\{ \pi ^{+}(z;1,s=q^{2n})\otimes 1\right\} \mathfrak{L}%
|_{W_{<n}}=(1\otimes q^{-\frac{n}{2}\sigma ^{z}})\left\{ \pi ^{(n-1)}\otimes
1\right\} L(zq^{n})(1\otimes q^{n\sigma ^{z}})
\end{equation}%
and%
\begin{equation}
\left\{ \pi ^{+}(z;1,s=q^{2n})\otimes 1\right\} \mathfrak{L}|_{W_{\geq
n}}=(1\otimes q^{-n\sigma ^{z}})\left\{ \pi ^{+}(zq^{2n};1,s=q^{-2n})\otimes
1\right\} \mathfrak{L}(1\otimes q^{2n\sigma ^{z}})\ .
\end{equation}%
Employing these identities together with the invariance of $W_{\geq n}$ one
can split the trace in the definition of the auxiliary matrix into two parts
leading to%
\begin{equation}
Q_{\alpha }(z;s=q^{2n})=q^{n(S^{z}-\alpha )}T_{\alpha
}^{(n-1)}(zq^{n})+q^{2n(S^{z}-\alpha )}Q_{\alpha }(zq^{2n};s=q^{-2n})\ .
\label{TnQQ2}
\end{equation}%
Apart from the factorization of $Q_{\alpha }$ into $Q_{\alpha }^{\pm }$ this
corresponds to the identity (\ref{TnQQ}). As it turns out this result is
already sufficient for our present purposes, i.e. the analytic continuation
of the fusion hierarchy in the spin-variable. See equation (\ref{defT2})
below.

\section{The trace function}

In order to make contact with the trace function used in \cite{BJMST} we now
reparametrize the fusion hierarchy and the $Q$-operator and take $M$ to be
even.

\subsection{Reparametrization}

Henceforth we set $z=q^{2\lambda },\ \zeta _{m}=q^{-\lambda _{m}}$ and
define the rescaled fusion hierarchy as%
\begin{equation}
T_{\alpha }^{(n)}(z)\rightarrow t_{\alpha }^{(n)}(\lambda ):=\frac{(zq)^{-%
\frac{M}{2}}T_{\alpha }^{(n)}(z)}{(q-q^{-1})^{M}}\tprod_{m}\zeta _{m}^{-1}\
\ .
\end{equation}%
This renormalization corresponds to the following choice of the six-vertex $%
R $-matrix which is in accordance with the conventions used in \cite{BJMST},%
\begin{equation}
r(\lambda )=\left( 
\begin{array}{cccc}
\lbrack \lambda +1]_{q} & 0 & 0 & 0 \\ 
0 & [\lambda ]_{q} & 1 & 0 \\ 
0 & 1 & [\lambda ]_{q} & 0 \\ 
0 & 0 & 0 & [\lambda +1]_{q}%
\end{array}%
\right) \ .
\end{equation}%
Thus, the fusion hierarchy is now expressed as%
\begin{equation}
t_{\alpha }^{(n)}(\lambda +\tfrac{n+1}{2})t_{\alpha }(\lambda )=t_{\alpha
}^{(n+1)}(\lambda +\tfrac{n}{2})\prod_{m}\left[ \lambda -\lambda _{m}+1%
\right] _{q}+t_{\alpha }^{(n-1)}(\lambda +\tfrac{n+2}{2})\prod_{m}[\lambda
-\lambda _{m}]_{q}
\end{equation}%
with the quantum determinant being%
\begin{equation}
t_{\alpha }^{(0)}(\lambda )=t^{(0)}(\lambda )=\prod_{m}[\lambda -\lambda
_{m}+\tfrac{1}{2}]_{q}\ .
\end{equation}%
Let us now turn to the re-definition of the auxiliary matrix. With respect
to the decomposition%
\begin{equation*}
\mathfrak{L}(\lambda )=\left( 
\begin{array}{cc}
\mathfrak{L}_{+}^{+} & \mathfrak{L}_{-}^{+} \\ 
\mathfrak{L}_{+}^{-} & \mathfrak{L}_{-}^{-}%
\end{array}%
\right)
\end{equation*}%
the matrix entries are now chosen as 
\begin{eqnarray}
\mathfrak{L}_{+}^{+} &=&\zeta ^{-\frac{1}{2}}~\frac{\zeta ^{2}/r~q^{\lambda +%
\frac{h_{1}+1}{2}}-q^{-\lambda -\frac{h_{1}+1}{2}}}{q-q^{-1}},\qquad s=\zeta
^{2},\;z=q^{2\lambda }  \notag \\
\mathfrak{L}_{-}^{+} &=&\zeta ^{-\frac{1}{2}}~q^{-\lambda }e_{0}q^{-\frac{%
h_{0}+1}{2}},  \notag \\
\mathfrak{L}_{+}^{-} &=&\zeta ^{-\frac{1}{2}}~q^{-\lambda }q^{-\frac{h_{1}-1%
}{2}}e_{1},  \notag \\
\mathfrak{L}_{-}^{-} &=&\zeta ^{-\frac{1}{2}}~\frac{r~q^{\lambda -\frac{%
h_{1}-1}{2}}-q^{-\lambda +\frac{h_{1}-1}{2}}}{q-q^{-1}}\ .
\end{eqnarray}%
Notice that this re-definition corresponds to the overall scaling factor%
\begin{equation}
Q_{\alpha }(z;r,s=\zeta ^{2})\rightarrow Q_{\alpha }(\lambda ;r,\zeta ):=%
\frac{(zq\zeta )^{-\frac{M}{2}}}{(q-q^{-1})^{M}}~Q_{\alpha }(z;r,s=\zeta
^{2})~\tprod_{m}\zeta _{m}^{-1}
\end{equation}%
and the eigenvalues of the auxiliary matrix in each spin-sector have now the
following decomposition%
\begin{equation}
Q_{\alpha }(\lambda ;r,\zeta =q^{\lambda ^{\prime }})=\mathfrak{N}_{\alpha
}~r^{\alpha -S^{z}}q^{\lambda ^{\prime }S^{z}}Q_{\alpha }^{+}(\lambda
)Q_{\alpha }^{-}(\lambda +\lambda ^{\prime })
\end{equation}%
with%
\begin{equation*}
\mathfrak{N}_{\alpha }=\left\{ 
\begin{array}{cc}
q^{-N^{\prime }(S^{z}+\alpha )}\tfrac{\left[ N^{\prime }(\alpha +S^{z})%
\right] _{q}}{\left[ \alpha +S^{z}\right] _{q}}~, & \text{if}\;q^{N}=1 \\ 
\frac{1}{q^{\alpha +S^{z}}-q^{-S^{z}-\alpha }}~, & \text{if }q\text{ generic}%
\end{array}%
\right.
\end{equation*}%
and%
\begin{equation}
Q_{\alpha }^{\pm }(\lambda )=\prod_{i=1}^{n_{\pm }}\left[ \lambda -\xi
_{i}^{\pm }\right] _{q},\qquad x_{i}^{\pm }=q^{-2\xi _{i}^{\pm }}\ .
\end{equation}%
Here we have used the sum rule 
\begin{equation*}
q^{-M}\tprod_{i=1}^{n_{+}}x_{i}^{+}\tprod_{i=1}^{n_{-}}x_{i}^{-}\tprod_{m}%
\zeta _{m}^{-2}=\frac{Q_{-\alpha }(0,q^{-1})}{Q_{\alpha }(0,q)}=\frac{%
q^{S^{z}-\alpha }-q^{\alpha -S^{z}}}{q^{-S^{z}-\alpha }-q^{\alpha +S^{z}}}
\end{equation*}%
which follows from combining (\ref{RQR1}) with (\ref{RQR2}) and setting $z=0$%
. Employing this decomposition the $TQ$ equation is equivalent to 
\begin{equation}
t_{\alpha }(\lambda )Q_{\alpha }^{\pm }(\lambda )=q^{-\alpha }Q_{\alpha
}^{\pm }(\lambda +1)\prod_{m=1}^{M}[\lambda -\lambda _{m}]_{q}+q^{\alpha
}Q_{\alpha }^{\pm }(\lambda -1)\prod_{m=1}^{M}[\lambda -\lambda _{m}+1]_{q}\
.
\end{equation}%
The expression for the transfer matrices of the fusion hierarchy is now%
\begin{equation}
t_{\alpha }^{(n-1)}(\lambda )=\frac{q^{-n\alpha }Q_{\alpha }^{+}(\lambda +%
\frac{n}{2})Q_{\alpha }^{-}(\lambda -\frac{n}{2})-q^{n\alpha }Q_{\alpha
}^{+}(\lambda -\frac{n}{2})Q_{\alpha }^{-}(\lambda +\frac{n}{2})}{%
q^{-S^{z}-\alpha }-q^{\alpha +S^{z}}},\quad n\in \mathbb{N}\ .  \label{tnQQ}
\end{equation}%
It is the last expression respectively (\ref{TnQQ}) which we want to
analytically continue in $n$ in order to obtain the trace-function used in
the description of the correlation functions of the infinite XXZ-chain.

\subsection{Analytic continuation and the limit $\protect\alpha \rightarrow
0 $}

As mentioned in the introduction we now analytically continue the expression
for the fusion hierarchy (\ref{TnQQ}), (\ref{TnQQ2}) in the spin-variable $%
n/2$ by defining the following operator%
\begin{eqnarray}
\boldsymbol{T}_{\alpha }(z,\zeta ) &=&(-)^{M}\frac{\zeta ^{S^{z}-\alpha
}Q_{\alpha }^{+}(z\zeta )Q_{\alpha }^{-}(z\zeta ^{-1})-\zeta ^{\alpha
-S^{z}}Q_{\alpha }^{+}(z\zeta ^{-1})Q_{\alpha }^{-}(z\zeta )}{%
q^{S^{z}-\alpha }-q^{\alpha -S^{z}}}  \label{defT} \\
&=&\frac{\zeta ^{\alpha -S^{z}}Q_{\alpha }(z\zeta ^{-1};s=\zeta ^{2})-\zeta
^{S^{z}-\alpha }Q_{\alpha }(z\zeta ;s=\zeta ^{-2})}{q^{S^{z}-\alpha
}-q^{\alpha -S^{z}}}~Q_{\alpha }(0)^{-1}\ .  \label{defT2}
\end{eqnarray}%
Note that this analytic continuation is unambiguous as the operators on the
right hand side of (\ref{defT2}) have only a polynomial dependence on the
spectral parameter $z$ by construction. The rescaled counterpart of (\ref%
{defT2}) gives the result (\ref{result}) stated in the introduction,%
\begin{equation}
\boldsymbol{t}_{\alpha }(\lambda ,\zeta )\equiv \frac{\boldsymbol{T}_{\alpha
}(z=q^{2\lambda },\zeta )}{q^{M(\lambda +1/2)}(q-q^{-1})^{M}}%
~\tprod_{m}\zeta _{m}^{-1}\ .  \label{deft}
\end{equation}%
First note that we recover the fusion hierarchy (\ref{TnQQ}) respectively (%
\ref{TnQQ2}) when setting $\zeta =q^{n},$%
\begin{equation}
\boldsymbol{T}_{\alpha }(z,\zeta =q^{n})=T_{\alpha }^{(n-1)}(z)\ .
\end{equation}%
Now, as an easy example for the occurrence of terms linear in $\lambda $ in
the matrix elements we evaluate (\ref{defT}) at the origin $z=0$. Then we
have for nonzero spin $S^{z}\neq 0$%
\begin{equation}
(-)^{M}\lim_{\alpha \rightarrow 0}\boldsymbol{T}_{\alpha }(0,\zeta
)=\lim_{\alpha \rightarrow 0}\frac{\zeta ^{-\alpha +S^{z}}-\zeta ^{\alpha
-S^{z}}}{q^{-\alpha +S^{z}}-q^{\alpha -S^{z}}}=\frac{\zeta ^{S^{z}}-\zeta
^{-S^{z}}}{q^{S^{z}}-q^{-S^{z}}}\ .  \label{defTr}
\end{equation}%
For vanishing spin $S^{z}=0$ we set $\zeta =q^{\lambda }$ and obtain 
\begin{equation}
(-)^{M}\lim_{\alpha \rightarrow 0}T_{\alpha }(0,\zeta =q^{\lambda })=\lambda
\ .  \label{dim}
\end{equation}%
These last two relations correspond to the defining equations (5.3) in \cite%
{BJMST}, see also (\ref{special}) in the introduction of this article. The
occurrence of linear terms $\lambda $ in the matrix elements is not
restricted to the zero spin sector but occurs more generally. To see this we
now derive the analogue of Lemma 5.1 in \cite{BJMST}. \medskip

\textbf{Lemma.} \emph{For }$M$ even \emph{the analytically continued fusion
hierarchy (\ref{defT}) decomposes in the limit }$\alpha \rightarrow 0$\emph{%
\ into a sum}%
\begin{equation}
\lim_{\alpha \rightarrow 0}\boldsymbol{T}_{\alpha }(z,\zeta =q^{\lambda
})=\lambda \cdot \boldsymbol{\tilde{G}}(z)+\boldsymbol{G}(z)
\end{equation}%
\emph{where the operators }$\boldsymbol{\tilde{G}},\boldsymbol{G}$\emph{\
are polynomial in the spectral variable }$z$\emph{\ and in each fixed spin
sector }$S^{z}\neq 0$\ \emph{have degrees}%
\begin{equation}
\deg \boldsymbol{\tilde{G}}|_{S^{z}}=M-|S^{z}|\qquad \text{\emph{and}}\qquad
\deg \boldsymbol{G}|_{S^{z}}=M\ .
\end{equation}%
\emph{If }$S^{z}=0$ \emph{then }%
\begin{equation}
\deg \boldsymbol{\tilde{G}}|_{S^{z}=0}=M\qquad \emph{and\qquad }\deg 
\boldsymbol{G}|_{S^{z}}<M\ .
\end{equation}%
\emph{According to the rescaling (\ref{deft}) this obviously implies (\ref%
{result2b}).}\medskip

\textbf{Remark 1.} For comparison with Lemma 5.1 in \cite{BJMST} we have to
identify $z\rightarrow \zeta _{1}^{2}$\ and $M=2p-2$, compare with (\ref%
{goal}) in this article. Moreover, we do consider here the degree of an
operator in a whole spin-sector not a single matrix element as in \cite%
{BJMST}.\medskip

\textbf{Remark 2.} Notice that for $M$ odd the linear terms in $\lambda $
are absent, since the terms containing $\limfunc{Tr}_{\pi ^{+}}q^{\alpha
h_{1}}$ which contains a pole in the limit $\alpha \rightarrow 0$ can never
occur in a matrix element of the $Q$-operator. See the proof below for an
explanation.

Note further, at roots of unity and periodic boundary conditions $\alpha =0$
the linear terms have not been observed in \cite{KQ4} and \cite{FM1,FM2}.
This is explained by the fact that the root of unity limit does not commute
with the limit $\alpha \rightarrow 0$. As long as $\alpha \neq 0$ (and
generic) the root of unity symmetries discussed in \cite{DFM,FM1,KQ4,FM3}
are not present, whence we arrive here at a different result.\medskip

\textsc{Proof.} Since we are only interested in determining the maximal
degree of the respective polynomials in $z$ we can set $\zeta _{m}=1$
without loss of generality. As an additional preparatory step we need to
make contact with the Casimir operator which has also been used to define
the trace function in \cite{BJMST}. Set $r=1$ in (\ref{piplus}), then we
have 
\begin{equation}
\lbrack e_{1},e_{0}]=z~\frac{sq^{h_{1}}-q^{-h_{1}}}{q-q^{-1}}  \label{comm1}
\end{equation}%
and 
\begin{equation}
\mathfrak{C}=z~\frac{sq^{h_{1}-1}+q^{-h_{1}+1}}{(q-q^{-1})^{2}}+e_{1}e_{0}=z~%
\frac{1+s}{(q-q^{-1})^{2}}\ .  \label{Cas2}
\end{equation}%
This identity corresponds to equation (5.4) in \cite{BJMST}, see (\ref%
{special}) in this article. Now consider a general matrix element of the
auxiliary matrix%
\begin{equation}
Q_{\alpha }(z;s)_{\underline{\varepsilon }}^{\underline{\varepsilon }%
^{\prime }}=\limfunc{Tr}_{\pi ^{+}}q^{\alpha h_{1}\otimes 1}\mathfrak{L}%
_{\varepsilon _{M}}^{\varepsilon _{M}^{\prime }}\cdots \mathfrak{L}%
_{\varepsilon _{1}}^{\varepsilon _{1}^{\prime }},\quad \tsum_{m}\varepsilon
_{m}=\tsum_{m}\varepsilon _{m}^{\prime }=2S^{z}
\end{equation}%
Since the $Q$-operator preserves the total spin the matrix elements $%
\mathfrak{L}_{\mp }^{\pm }$ always occur in pairs and we compute%
\begin{eqnarray}
\limfunc{Tr}_{\pi ^{+}}\{X~\mathfrak{L}_{\mp }^{\pm }\mathfrak{L}_{\pm
}^{\mp }\} &=&zq~\limfunc{Tr}_{\pi ^{+}}\{X~(q-q^{-1})^{2}\mathfrak{C}%
-sq^{h_{1}\pm 1}-q^{-h_{1}\mp 1}\}  \notag \\
&=&zq~\limfunc{Tr}_{\pi ^{+}}\{X~\left( 1+s-q^{-h_{1}\mp 1}-sq^{h_{1}\pm
1}\right) \}
\end{eqnarray}%
The numbers of the various operators $\mathfrak{L}_{\varepsilon
}^{\varepsilon ^{\prime }}$ occurring in a matrix element is given by%
\begin{equation}
\#\mathfrak{L}_{\pm }^{\pm }+\#\mathfrak{L}_{\pm }^{\mp }=\frac{M}{2}\pm
S^{z}=n_{\mp }\quad \text{and\quad }\#\mathfrak{L}_{-}^{+}=\#\mathfrak{L}%
_{+}^{-}=\tsum_{m}\frac{1-\varepsilon _{m}\varepsilon _{m}^{\prime }}{2}=:n_{%
\underline{\varepsilon },\underline{\varepsilon }^{\prime }}\ .
\end{equation}%
Using the above relations any matrix element of the $Q$-operator in a fixed
spin sector can be expressed as a linear combination of terms of the form%
\begin{multline*}
Q_{\alpha }(z\zeta ;s=\zeta ^{-2})_{\underline{\varepsilon }}^{\underline{%
\varepsilon }^{\prime }}=(zq)^{n_{\underline{\varepsilon },\underline{%
\varepsilon }^{\prime }}}\sum c_{\underline{\varepsilon },\underline{%
\varepsilon }^{\prime }}~\limfunc{Tr}_{\pi ^{+}}\left\{ q^{\alpha
h_{1}}\left( \zeta ^{-1}~zq^{1+\frac{h_{1}}{2}}-q^{-\frac{h_{1}}{2}}\right)
^{n_{-}-n_{\underline{\varepsilon },\underline{\varepsilon }^{\prime
}}}\right. \\
\times \left. \left( \zeta ~zq^{1-\frac{h_{1}}{2}}-q^{\frac{h_{1}}{2}%
}\right) ^{n_{+}-n_{\underline{\varepsilon },\underline{\varepsilon }%
^{\prime }}}\tprod_{i=1}^{n_{\underline{\varepsilon },\underline{\varepsilon 
}^{\prime }}}\zeta ^{\sigma _{i}}q^{-\sigma _{i}h_{1}}\right\}
\end{multline*}%
where the coefficients $c_{\underline{\varepsilon },\underline{\varepsilon }%
^{\prime }}$ do not depend on $z$ and $\sigma _{i}=0,\pm 1$ can vary in each
factor, but there can be terms present for which all the $\sigma _{i}$'s are
equal. To see this note that $[\mathfrak{L}_{+}^{+},\mathfrak{L}_{-}^{-}]=0$%
, while the commutation of $\mathfrak{L}_{\pm }^{\pm }$ with $\mathfrak{L}%
_{\pm }^{\mp }$ only produces powers in $q$. The operators $\mathfrak{L}%
_{-}^{+}$ and $\mathfrak{L}_{+}^{-}$ commute according to (\ref{comm1}). Let
us distinguish the two cases $n_{\underline{\varepsilon },\underline{%
\varepsilon }^{\prime }}=0$ and $n_{\underline{\varepsilon },\underline{%
\varepsilon }^{\prime }}\neq 0$.

Choose a matrix element with $n_{\underline{\varepsilon },\underline{%
\varepsilon }^{\prime }}=0$, i.e. $\varepsilon _{m}=\varepsilon _{m}^{\prime
}$. This is possible in all spin-sectors. Then the term of maximal degree
has the coefficient%
\begin{multline*}
Q_{\alpha }(z\zeta ;s=\zeta ^{-2})_{\underline{\varepsilon }}^{\underline{%
\varepsilon }}= \\
\limfunc{Tr}_{\pi ^{+}}\left\{ q^{\alpha h_{1}}\left( \zeta ^{-1}~zq^{1+%
\frac{h_{1}}{2}}-q^{-\frac{h_{1}}{2}}\right) ^{\frac{M}{2}-|S^{z}|}\left(
\zeta ~zq^{1-\frac{h_{1}}{2}}-q^{\frac{h_{1}}{2}}\right) ^{\frac{M}{2}%
-|S^{z}|}\right. \\
\times \left. \left( \zeta ^{-\sigma }zq^{1+\sigma \frac{h_{1}}{2}%
}-q^{-\sigma \frac{h_{1}}{2}}\right) ^{2|S^{z}|}\right\} =(zq)^{M}\zeta
^{-2S^{z}}\limfunc{Tr}_{\pi ^{+}}q^{\alpha (1+S^{z})h_{1}}+~...
\end{multline*}%
Here we have set $\sigma =\limfunc{sgn}S^{z}$. From this we infer using (\ref%
{defT2}) that%
\begin{multline*}
\frac{\zeta ^{S^{z}-\alpha }Q_{\alpha }(z\zeta ;s=\zeta ^{-2})-\zeta
^{\alpha -S^{z}}Q_{\alpha }(z\zeta ^{-1};s=\zeta ^{2})}{(q^{S^{z}-\alpha
}-q^{\alpha -S^{z}})Q_{\alpha }(0)}= \\
(zq)^{M}\frac{\zeta ^{-S^{z}-\alpha }-\zeta ^{\alpha +S^{z}}}{%
(q^{S^{z}-\alpha }-q^{\alpha -S^{z}})Q_{\alpha }(0)}\limfunc{Tr}_{\pi
^{+}}q^{\alpha (1+S^{z})h_{1}}+~...
\end{multline*}%
Setting $\zeta =q^{\lambda }$ we conclude similarly to our previous
calculation at $z=0$ that the linear terms in $\lambda $ originate from
coefficients containing $\limfunc{Tr}_{\pi ^{+}}q^{\alpha h_{1}}$ which
develops a pole in the limit $\alpha \rightarrow 0$. Thus, we deduce that $%
\deg \boldsymbol{\tilde{G}}=M$ and $\deg \boldsymbol{G}<M$ when $S^{z}=0$.
When $S^{z}\neq 0,$ on the other hand, we have as degrees $\deg \boldsymbol{G%
}=M$ and $\deg \boldsymbol{\tilde{G}}=M-|S^{z}|$. Note that for this it is
crucial that $2|S^{z}|$ is an even integer which is only the case when $M$
is even.

Now choose a matrix element with $n_{\underline{\varepsilon },\underline{%
\varepsilon }^{\prime }}\neq 0$. This is always possible as long as $%
|S^{z}|<M/2$. All we need to show is that the just derived degrees are not
exceeded. The term of maximal degree has now coefficients of the form%
\begin{equation*}
Q_{\alpha }(z\zeta ;s=\zeta ^{-2})_{\underline{\varepsilon }}^{\underline{%
\varepsilon }^{\prime }}=(zq)^{M-n_{\underline{\varepsilon },\underline{%
\varepsilon }^{\prime }}}\sum c_{\underline{\varepsilon },\underline{%
\varepsilon }^{\prime }}\zeta ^{-2S^{z}}\limfunc{Tr}_{\pi ^{+}}q^{(\alpha
+S^{z})h_{1}}\tprod_{i=1}^{n_{\underline{\varepsilon },\underline{%
\varepsilon }^{\prime }}}\zeta ^{\sigma _{i}}q^{-\sigma _{i}h_{1}}+~...
\end{equation*}%
As in the earlier examples we need to determine the maximal degree of the
term which contains $\limfunc{Tr}_{\pi ^{+}}q^{\alpha h_{1}}$ yielding the
linear dependence on $\lambda $ in the limit $\alpha \rightarrow 0$. The
degree of $\boldsymbol{\tilde{G}}$ is maximized by choosing a matrix element
with $n_{\underline{\varepsilon },\underline{\varepsilon }^{\prime
}}=|S^{z}| $, where the coefficient of the term with all $\sigma _{i}=%
\limfunc{sgn}S^{z} $ is non-vanishing. Then we have 
\begin{equation*}
\zeta ^{S^{z}-\alpha }Q_{\alpha }(z\zeta ;\zeta ^{-2})_{\underline{%
\varepsilon }}^{\underline{\varepsilon }^{\prime }}-\zeta ^{\alpha
-S^{z}}Q_{\alpha }(z\zeta ^{-1};\zeta ^{2})_{\underline{\varepsilon }}^{%
\underline{\varepsilon }^{\prime }}=\text{const.}~(zq)^{M-|S^{z}|}(\zeta
^{-\alpha }-\zeta ^{\alpha })\limfunc{Tr}_{\pi ^{+}}q^{\alpha h_{1}}+~...
\end{equation*}%
from which we infer in the limit $\alpha \rightarrow 0$ that the maximal
degree of $\boldsymbol{\tilde{G}}$\ in a fixed spin sector is again $%
M-|S^{z}|$. For the remainder polynomial $\boldsymbol{G}$ we find as before
that its degree is strictly smaller than $M$ if $S^{z}=0$. $_{\square }$

\subsubsection{Example: $M=4$ and $S^{z}=1$}

Let us consider a simple example for the homogeneous chain to show how the
linear term in $\lambda $ emerges in a matrix element of the operator (\ref%
{defT}). Setting $M=4$ and $S^{z}=1$ we choose the matrix element%
\begin{eqnarray}
Q_{\alpha }(z;s)_{+-++}^{-+++} &=&\limfunc{Tr}_{\pi ^{+}}q^{\alpha
h_{1}}(zsq^{1+\frac{h_{1}}{2}}-q^{-\frac{h_{1}}{2}})^{2}zq\left(
1+s-q^{-h_{1}-1}-sq^{h_{1}+1}\right)  \notag \\
&=&zq\limfunc{Tr}_{\pi ^{+}}q^{\alpha
h_{1}}(z^{2}s^{2}q^{2+h_{1}}-2zsq+q^{-h_{1}})\left(
1+s-q^{-h_{1}-1}-sq^{h_{1}+1}\right)  \notag \\
&=&-zsq^{2}\{z^{2}s+1+2z(1+s)\}\limfunc{Tr}_{\pi ^{+}}q^{\alpha h_{1}}+~...
\end{eqnarray}%
In the last line we have only written out the terms which will give rise to
the linear dependence in $\lambda $. Namely, inserting this expression into (%
\ref{defT}) we find%
\begin{eqnarray}
\boldsymbol{T}_{\alpha }(z;\zeta )_{+-++}^{-+++} &=&\frac{\zeta ^{1-\alpha
}Q_{\alpha }(z\zeta ;\zeta ^{-2})-\zeta ^{\alpha -1}Q_{\alpha }(z\zeta
^{-1};\zeta ^{2})}{Q_{\alpha }(0)(q^{1-\alpha }-q^{\alpha -1})}  \notag \\
&=&-zq^{2}(z^{2}+1+2z(\zeta +\zeta ^{-1}))~\frac{\zeta ^{-\alpha }-\zeta
^{\alpha }}{Q_{\alpha }(0)(q^{1-\alpha }-q^{\alpha -1})}~\limfunc{Tr}_{\pi
^{+}}q^{\alpha h_{1}}+~...  \notag \\
&=&-zq^{2}(z^{2}+1)\frac{\zeta ^{-\alpha }-\zeta ^{\alpha }}{Q_{\alpha
}(0)(q^{1-\alpha }-q^{\alpha -1})}~\limfunc{Tr}_{\pi ^{+}}q^{\alpha
h_{1}}+~...
\end{eqnarray}%
Let us distinguish the case when $q$ is generic and when it is a root of
unity. If $q^{N}=1$ then%
\begin{equation}
Q_{\alpha }(0)=\frac{1-q^{-2N^{\prime }\alpha }}{q^{\alpha -1}-q^{1-\alpha }}%
\qquad \text{and}\qquad \limfunc{Tr}_{\pi ^{+}}q^{\alpha h_{1}}=\frac{%
1-q^{-2N^{\prime }\alpha }}{q^{\alpha }-q^{-\alpha }}\ .
\end{equation}%
On the other hand we have for generic $q$ 
\begin{equation}
Q_{\alpha }(0)=\frac{1}{q^{\alpha -1}-q^{1-\alpha }}\qquad \text{and}\qquad 
\limfunc{Tr}_{\pi ^{+}}q^{\alpha h_{1}}=\frac{1}{q^{\alpha }-q^{-\alpha }}\ ,
\end{equation}%
where these expressions are understood as analytic continuation from the
region where the trace converges. Thus, setting $\zeta =q^{\lambda }$ we
arrive at%
\begin{eqnarray}
\lim_{\alpha \rightarrow 0}\boldsymbol{T}_{\alpha }(z;\zeta )_{+-++}^{-+++}
&=&-zq^{2}(z^{2}+1+2z(q^{\lambda }+q^{-\lambda }))~\lim_{\alpha \rightarrow
0}\frac{q^{\alpha \lambda }-q^{-\alpha \lambda }}{q^{\alpha }-q^{-\alpha }}%
+\ ...\   \notag \\
&=&-\lambda ~zq^{2}(z^{2}+1+2z(q^{\lambda }+q^{-\lambda }))+~...\ 
\end{eqnarray}%
where the coefficient of $\lambda $ is of degree $M-|S^{z}|=3$ in $z$ in
accordance with our lemma.

\section{Conclusions}

In this paper we continued a previous study \cite{KQ,KQ2,KQ3,KQ4,KQ5}\ on
the explicit construction of operator-solutions to Baxter's TQ equation (\ref%
{TQ}). In terms of eigenvalues this equation is a second order difference
equation and its theory resembles closely the one of second order ordinary
differential equations. In the case of the XXZ chain we discussed the
existence of two linearly independent solutions when quasi-periodic boundary
conditions are imposed by explicitly constructing the relevant $Q$-operators
using representation theory. Because it is usually required that $Q$%
-operators should commute for arbitrary values of the spectral parameter, $%
[Q(z),Q(w)]=0$, this precludes by choice of the construction method the
possibility of obtaining \textquotedblleft non-analytic\textquotedblright\
solutions to the $TQ$ equation: that is, solutions which obtain the linear
terms discussed in the text and which have been postulated in \cite{PrSt}
for even chains with periodic boundary conditions and when $q$ is generic.
The result of this paper is that such solutions can arise by taking the
limit from quasi-periodic to periodic boundary conditions in the explicitly
constructed $Q$-operators. Recall that for generic $q$ this limit required a
careful analysis. First one had to choose the twist parameter $\alpha $ such
that convergence of the trace over the infinite-dimensional auxiliary space
is guaranteed. In a second step we then analytically continued the matrix
elements in $\alpha $ from the region of convergence to the complex plane
which enabled us in the final step to discuss the limit $\alpha \rightarrow
0 $. To complete the investigation by computing the spectra of the $Q$%
-operators in this limit one would need to know the explicit dependence of
the Bethe roots on the twist parameter. This is a rather formidable
challenge as the solutions to the Bethe ansatz equations are in general not
known.

The main motivation for our construction of this \textquotedblleft
non-analytic\textquotedblright\ $Q$-operator has been the relation with the
recent developments in the computation of correlation functions as explained
in detail in the introduction. The alternative expression (\ref{result}) for
the special trace of the monodromy matrix (\ref{goal}) entering the ansatz
in \cite{BJMST} shows that there is a more fundamental quantity in which the
correlation functions can be expressed and provides a different point of
view on the role of the trace function (\ref{defTr}). In future work it
needs to be explored whether there are concrete practical implications of
the identity (\ref{result}) which facilitate the computation of correlation
functions. For instance, one might ask whether one can insert each of the
two terms in the difference (\ref{result}) separately in the ansatz for the
correlation functions and if they satisfy identities analogous to the
quantum Knizhnik-Zamolodchikov equations. Of course one has to keep in mind
that in order to perform the limit $\alpha \rightarrow 0$ both terms will be
needed at the end. However, such an investigation might yield further
insight into the analytic structure.{\small \medskip }

\textbf{Note added in Proof}: After this paper was completed the work \cite%
{BJMSTXYZ} appeared which extends the investigation of correlation functions
to the inhomogeneous eight-vertex or XYZ model. An analogous $Q$-operator
for this model would be helpful as it would simplify the computation of the
analogue of the trace function over the Sklyanin algebra. The present
discussion provides a first step towards this aim.{\small \medskip }

\noindent \textbf{Acknowledgments}. The author wishes to thank Professor
Michio Jimbo for his kind invitation and the suggestion of the problem
discussed in this article. He also would like to express his gratitude for
the hospitality of the Graduate School of Mathematical Science, Unviversity
of Tokyo where this work has been carried out. This project is part of a
University Research Fellowship of the Royal Society.

\appendix

\end{document}